\documentclass[a4paper,preprint]{emulateapj}
\usepackage{amstext}
\usepackage{apjfonts}
\usepackage{amsmath}
\newcommand{\bfig}{\noindent\begin{minipage}{3.48in}}
\newcommand{\efig}{\bigskip\end{minipage}}

\begin{document}

\title{Linear Polarization in Gamma-Ray Bursts: The Case for an Ordered Magnetic Field}


\author{Jonathan Granot\altaffilmark{1} and Arieh K\"onigl\altaffilmark{2}}
\altaffiltext{1}{Institute for Advanced Study, Olden Lane,
Princeton, NJ 08540; granot@ias.edu} \altaffiltext{2}{Department
of Astronomy \& Astrophysics and Enrico Fermi Institute,
University of Chicago, 5640 S. Ellis Ave., Chicago, IL 60637;
arieh@jets.uchicago.edu}

\begin{abstract}
  
  Linear polarization at the level of $\sim 1-3\%$ has by now been
  measured in several GRB afterglows. Whereas the degree of
  polarization, $P$, was found to vary in some sources, the position
  angle, $\theta_p$, was roughly constant in all cases.  Until now,
  the polarization has been commonly attributed to synchrotron
  radiation from a jet with a tangled magnetic field that is viewed
  somewhat off axis. However, this model predicts either a peak in $P$
  or a $90^\circ$ change in $\theta_p$ around the ``jet break'' time
  in the lightcurve, for which there has so far been no observational
  confirmation. We propose an alternative interpretation, wherein the
  polarization is attributed, at least in part, to a large-scale,
  ordered magnetic field in the ambient medium.  The ordered component
  may dominate the polarization even if the total emissivity is
  dominated by a tangled field generated by postshock turbulence. In
  this picture, $\theta_p$ is roughly constant because of the
  uniformity of the field, whereas $P$ varies as a result of changes
  in the ratio of the ordered-to-random mean-squared field amplitudes.
  We point out that variable afterglow light curves should be
  accompanied by a variable polarization. The radiation from the
  original ejecta, which includes the prompt $\gamma$-ray emission and
  the emission from the reverse shock (the `optical flash' and `radio
  flare'), could potentially exhibit a high degree of polarization (up
  to $\sim 60\%$) induced by an ordered transverse magnetic field
  advected from the central source.

\end{abstract}


\section{Introduction}
\label{introduction}
The first detection of polarization in an optical afterglow was in GRB
990510, where a degree of polarization $P=1.7\pm 0.2\%$ ($1.6\pm
0.2\%$) was measured at $t_{\rm obs} \approx 18\;$hr ($21\;$hr) after
the burst.\footnote{See Covino et al. 2003a for references to the
  above observations as well as to subsequent polarization
  measurements.}  Since then, $P\sim1-3\%$ has been detected in a few
additional afterglows,\footnote{There is one exception, $P=9.9\pm
  1.3\%$, measured in GRB 020405 at $t_{\rm obs}=1.3\;$days (Bersier
  et al. 2003).  Significantly lower values ($P\approx 1.5-2\%$) were
  measured in this afterglow at $t_{\rm obs}=1.2,2.2$, and $3.3\;$days
  by other groups, with a similar $\theta_p$. If real, this behavior
  has no simple explanation in any of the existing models.} some of
which showed a temporal variation in $P$ (Rol et al. 2000; Barth et
al. 2003), but typically the position angle (PA) $\theta_p$ showed
little or no change. A few other afterglows produced only upper
limits, $P\lesssim 2-5\%$.

The polarization is attributed to synchrotron emission behind a shock
wave. It thus depends on the local magnetic field configuration, which
determines the polarization at each point of the afterglow image, and
on the global geometry of the shock, which determines how the
polarization is averaged over the (unresolved) image. We make a
distinction between magnetic field configurations that are axially
symmetric about the normal $\hat{n}_{\rm sh}$ to the shock surface and
those that are not.\footnote{In this picture, the field is tangled
  over very small scales and possesses this symmetry when averaged
  over regions of angular size $\ll 1/\gamma$, where $\gamma$ is the
  Lorentz factor of the shocked fluid.}  The first category gives no
net polarization for a spherical flow and is assumed in most previous
works (Sari 1999; Ghisellini \& Lazzati 1999; Medvedev \& Loeb 1999;
Granot et al.  2002; Rossi et al. 2002). These models take the field
to be completely random in the plane of the shock, and the
polarization is usually attributed to a jet viewed somewhat off axis.
For a structured jet, $P$ has one peak near the jet break time $t_j$
(when $1/\gamma$ increases to $\sim \theta_0$, the initial jet opening
half-angle), whereas for a uniform jet $P$ has 2 (or 3) peaks near
$t_j$, with $P$ passing through zero and $\theta_p$ changing by
$90^\circ$ between the peaks.

The second category can produce net polarization even for a spherical
flow. One example is the ``patchy coherent field'' model of Gruzinov
\& Waxman (1999), where the observed region consists of $N\sim 50$
mutually incoherent patches of angular size $\lesssim 1/\gamma$,
within each of which the field is fully ordered.\footnote{A similar
  model was used to study the linear polarization induced by
  microlensing of GRB afterglows (Loeb \& Perna 1998).} This model
predicts $P\sim P_{\rm max}/N^{1/2}\sim 10\%$ (where $P_{\rm max}\sim
60-70\%$ is the maximum $P$ of local synchrotron emission in a uniform
magnetic field) and simultaneous (random) variability in $P$ and
$\theta_p$ on timescales $\Delta t_{\rm obs}\lesssim t_{\rm obs}$ .
The idea behind this model is that $\sim 1/\gamma$ is the angular size
of causally connected regions, and for a magnetic field that is
generated in the shock itself this is the largest scale over which the
field can be coherent.

However, if the magnetic field were ordered on an angular scale
$\theta_B\gtrsim 1/\gamma$, then the resulting $P$ could approach
$P_{\rm max}$.  Such a situation can be realized if an ordered field
exists in the medium into which the shock propagates.  For a typical
interstellar medium, the postshock field would still be very weak
(with the magnetic energy a fraction $\epsilon_B\lesssim 10^{-10}$ of
the internal energy), but it would be higher ($\epsilon_B\lesssim
10^{-4}$; Biermann \& Cassinelli 1993) if the shock expands into a
magnetized wind of a progenitor star, and higher yet ($\epsilon_B\sim
0.01-0.1$) if it propagates into a pulsar-wind bubble (PWB), as
expected in the supranova model (K\"onigl \& Granot 2002).  A strong
ordered field component is likely to exist in the original ejecta and
could give rise to a high value of $P$ ($\lesssim P_{\rm max}$) in
both the prompt GRB and the reverse-shock emission.

We calculate the polarization for a jet with a tangled magnetic field
in \S~\ref{B_rnd}, investigate the effects of adding an ordered field
component in \S~\ref{B_ord}, and discuss the results in
\S~\ref{discussion}.

\section{Polarization from a Jet with a Tangled Magnetic Field}
\label{B_rnd}

Synchrotron emission is generally partially linearly polarized. In
terms of the Stokes parameters: $V\equiv 0$,
$\theta_p=\frac{1}{2}\arctan(U/Q)$, and $P=(Q^2+U^2)^{1/2}/I$. As the
Stokes parameters are additive for incoherent emission, they can be
calculated by summing over all the contributions from different fluid
elements to the flux at a given observed time $t_{\rm obs}$. In
practice, the flux is calculated by dividing $t_{\rm obs}$ into bins
of size $\delta t_{\rm obs}$, and assigning to the appropriate time
bin the contribution to $F_{\nu}$ from the emission of each 4-volume
fluid element $d^4x$,
\begin{equation}\label{fnu}
dF_\nu(t_{\rm obs},\hat{n},{\bf r},t) = {(1+z)^2\over d_L^2}
{\,j'_{\nu'}({\bf r},t)
\Theta(\delta t_{\rm obs}-2|t-\frac{\hat{n}\cdot{\bf r}}{c}
-t_{\rm obs}|)d^4x \over \gamma^2({\bf r},t)
\left[ 1 - \hat{n}\cdot{\bf v}({\bf r},t)/c\right]^2\delta
t_{\rm obs}}\ ,
\end{equation}
\begin{equation}\label{QU}
\left\{\begin{matrix} U/I \vspace{0.12cm}\cr Q/I\end{matrix}\right\}
=\left(\sum dF_\nu\right)^{-1}\sum dF_\nu \left\{\begin{matrix} P\sin
2\theta_p\vspace{0.12cm}\cr P\cos 2\theta_p \end{matrix}\right\}\ ,
\end{equation}
where $j'_{\nu'}$ is the local emissivity, ${\bf v}$ is the fluid
velocity, $\hat{n}$ is the direction to the observer, $\Theta$ is the
Heaviside step function and the summation in Eq. \ref{QU} is over
${\bf r}$ \& $t$, for fixed $t_{\rm obs}$ \& $\hat{n}$.

In this section, we consider only a random magnetic field $B_{\rm
  rnd}$ that is tangled on angular scales $\ll 1/\gamma$, with axial
symmetry w.r.t. $\hat{n}_{\rm sh}$.  The field anisotropy is
parameterized by $b\equiv2\langle B_\parallel^2\rangle/\langle
B_\perp^2\rangle$, where $B_\parallel$ ($B_\perp$) is the magnetic
field component parallel (perpendicular) to $\hat{n}_{\rm sh}$.  The
local polarization of the emission from a given fluid element
is\footnote{For simplicity, we assume
  $j'_{\nu'}\propto(B'\sin\chi')^\epsilon$, where $\cos\chi'=
  \hat{n}'\cdot\hat{B}'$, with $\epsilon=2$ (Sari 1999), although in
  reality $\epsilon$ may be different.}
\begin{equation}\label{pol_rnd}
{P_{\rm rnd}(\theta')\over P_{\rm max}}= {\left(\langle
B_\parallel^2\rangle-\langle B_\perp^2\rangle/2\right)
\sin^2\theta' \over \langle B_\parallel^2\rangle\sin^2\theta'+
(1+\cos^2\theta')\langle B_\perp^2\rangle/2}=
{(b-1)\sin^2\theta'\over 2+(b-1)\sin^2\theta'}\ ,
\end{equation}
(Gruzinov 1999; Sari 1999), where $\cos\theta'=\frac{\mu-v/c}{1-\mu
  v/c}$, $\mu\equiv \hat{n}\cdot\hat{n}_{\rm sh}$.  For $P>0$, the
polarization is in the direction of $\hat{n}\times\hat{n}_{\rm sh}$.

The magnetic field configuration behind the shock, and hence the value
of $b$, cannot be easily deduced from first principles. It was
suggested that small-scale postshock fields can be generated by a
two-stream instability (e.g., Medvedev \& Loeb 1999), which predicts
$b\ll 1$. However, it is not clear whether the magnetic fields
produced in this way survive in the bulk of the postshock flow
(Gruzinov 1999, although see Frederiksen et al. 2003), or whether this
is the dominant tangling mechanism. The relatively low observed values
of $P$ ($\lesssim 3-4\%$) suggest that $0.5\lesssim b\lesssim 2$ if
the polarization is due to a jet with a shock-generated field.

Turbulence in the postshock region (possibly induced by a
microinstability) could amplify and isotropize the field, keeping $b$
close to $1$. As each fluid element moves downstream from the shock
transition, it is sheared by the flow. For a Blandford-McKee (1976)
self-similar blastwave solution with an ambient density $\rho_{\rm
  ext}\propto r^{-k}$, the length of a fluid element in the directions
parallel and perpendicular to $\hat{n}_{\rm sh}$ scales with the
self-similarity variable $\chi$ (where $\chi=1$ at the shock front and
increases with distance behind the shock) as
$L_\parallel\propto\chi^{\frac{9-2k}{2(4-k)}}$ and
$L_\perp\propto\chi^{\frac{1}{4-k}}$.  Therefore, the stretching of
each fluid element in the radial direction is larger than in the
tangential direction. This would increase $b$ while maintaining axial
symmetry about $\hat{n}_{\rm sh}$. The relevant value of $b$ here is
the average over the postshock region, weighted by the emissivity. If
the turbulence only persists over a small distance ($0<\chi-1\ll 1$,
but still $\gg$ the plasma skin depth) then, since most of the
emission is from $\chi\lesssim$ a few, we may have $1<b\lesssim\;$a
few.\footnote{On the other hand, if an intrinsically isotropic
  turbulence persists over a large portion of the emission region,
  this could reduce the effect of shearing on $b$, resulting in
  $0<(b-1)\ll 1$.}

\begin{figure}
\plotone{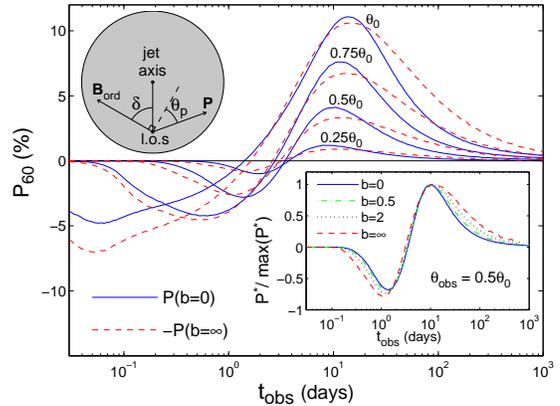}
\figcaption[]{\label{fig1}
Polarization lightcurves for a jetted GRB
afterglow with a random magnetic field. The solid (dashed) lines
are for $b=0$ ($b=\infty$, with a minus sign); $P_{60}\equiv
P(0.6/P_{\rm max})$. The jet parameters are the same as in Fig. 3
of Granot et al. (2002). The lower right inset shows
$P^{*}\equiv{\rm sgn}(1-b)P$, normalized to its maximum value, for
a viewing angle $\theta_{\rm obs}=0.5\,\theta_0$ and
$b=0,0.5,2,\infty$, making it easier to follow the effect of $b$
on the shape of the lightcurve. The upper left inset shows a
schematic diagram of the plane of the sky. The shaded region
represents a jet with both a tangled and ordered field components.
The projection of the ordered magnetic field on the plane of the
sky, ${\bf B}_{\rm ord}$, is at an angle $\delta$ in the
counterclockwise direction w.r.t. the direction from the line of
sight (l.o.s.) to the jet axis. The polarization vector ${\bf P}$
is at an angle $\theta_p$, measured clockwise from the
perpendicular to ${\bf B}_{\rm ord}$.}
\vspace{-0.5cm}
\end{figure}

Figure \ref{fig1} shows the polarization lightcurves for $b=0$ and
$\infty$, based on the jet model of Kumar \& Panaitescu (2000).  For
all viewing angles $\theta_{\rm obs}\leq\theta_0$ there are two peaks
in $P$, a little before and after $t_j$: $P$ passes through zero in
between these peaks as $\theta_p$ changes by $90^\circ$. This result
is similar to that of Ghisellini \& Lazzati (1999), who did not
consider lateral spreading of the jet, and differs from that of Sari
(1999), who assumed $\theta_{\rm jet}(t_{\rm obs}$$>$$t_j)=1/\gamma$,
since the lateral spreading in the jet model that we use is smaller
than the one used by Sari. The main distinction between the $b<1$ and
$b>1$ cases is a $90^\circ$ difference in $\theta_p$, but this
prediction can only be tested if one can independently determine the
direction from the l.o.s. to the jet axis. This may in principle be
done by measuring the direction of motion of the flux centroid
$\hat{n}_c$ (Sari 1999): for $b<1$, $\hat{P}$ is perpendicular to
(aligned with) $\hat{n}_c$ before (after) $t_j$, whereas for $b>1$ the
situation is reversed.  More generally, one would then be able to test
if indeed $\hat{P}\parallel\hat{n}_c$ or $\hat{P}\perp\hat{n}_c$, as
expected for a pure $B_{\rm rnd}$ field, or if the angle between them
is different, as would generally be the case if an ordered field
component were also present.

\section{The Effects of an Ordered Magnetic Field}
\label{B_ord}

We now add an ordered magnetic field component $B_{\rm ord}$ to the
random field $B_{\rm rnd}$ considered in \S~\ref{B_rnd}. In passing
through the shock transition, the parallel component of the ambient
magnetic field $B_{\rm ext}$ remains unchanged but the transverse
component is amplified by a factor equal to the fluid compression
ratio, which for $\gamma\gg 1$ is $4\gamma$. Thus typically
$B_\perp\gg B_\parallel$ behind the shock. For simplicity we assume
that $B_{\rm ord}$ lies in the plane of the shock and is fully ordered
and that $B_{\rm ext}$ is uniform, so that $B_{\rm ord}$ is coherent
over the entire shock.

It is most convenient to sum over the Stokes parameters associated
with $B_{\rm rnd}$ and $B_{\rm ord}$ separately and combine them at
the end.\footnote{This is valid in the limit where the two components
  are associated with distinct fluid elements.  Alternative schemes
  for combining $B_{\rm rnd}$ and $B_{\rm ord}$ may produce a somewhat
  different polarization.} The direction of polarization $\hat{P}_{\rm
  ord}$ of the emission from the ordered component is perpendicular to
its projection (${\bf B}_{\rm ord}$) on the plane of the sky;
$\hat{P}_{\rm rnd}$ is either along the plane containing the l.o.s.
and the jet symmetry axis (for $P_{\rm rnd}>0$) or perpendicular to
that direction (for $P_{\rm rnd}<0$).  The total polarization and the
PA are given by
\begin{eqnarray}\label{P_tot}
P &=& \left({\eta P_{\rm ord}\over 1+\eta}\right) \left [
1+\left({P_{\rm rnd}\over\eta P_{\rm ord}}\right)^2
-2\left({P_{\rm rnd}\over\eta P_{\rm ord}}\right)\cos
2\delta\right ]^{1/2}\ ,
\\ \label{theta_p}
\theta_p &=& {1\over 2}\arctan\left(
{\sin 2\delta\over\cos 2\delta-\eta P_{\rm ord}/P_{\rm rnd}}\right)\ ,
\end{eqnarray}
where $\eta\equiv I_{\rm ord}/I_{\rm rnd}\approx \langle B_{\rm
  ord}^2\rangle/\langle B_{\rm rnd}^2\rangle$ is the ratio of the
observed intensities in the two components and $\theta_p\,,\,\delta$
are measured as illustrated in the upper left inset of Figure
\ref{fig1}.

For $B_{\rm ord}$ we find that the PA as a function of the polar angle
$\theta$ from the l.o.s. and the azimuthal angle $\phi$ (measured from
$\hat{B}_{\rm ord}$) is given, in the relativistic ($\gamma\gg 1$)
limit, by $\theta_p=\phi+\arctan(\frac{1-y}{1+y}\cot\phi)$, where
$y\equiv(\gamma\theta)^2$. We have $I_\nu=I'_{\nu'}(\nu/\nu')^3$, with
$I'_{\nu'}\propto\nu'^{-\alpha}[1-(\hat{n}'\cdot\hat{B}'_{\rm
  ord})^2]^{\epsilon/2}$, where $\nu/\nu'\approx \frac{2\gamma}{1+y}$
and $1-(\hat{n}'\cdot\hat{B}'_{\rm ord})^2\approx
\left(\frac{1-y}{1+y}\right)^2\cos^2\phi+\sin^2\phi$. The Stokes
parameters are given by $(U,Q)/IP_{\rm max}=\int d\Omega I_\nu(\sin
2\theta_p,\cos 2\theta_p) /\int d\Omega I_\nu$. For a spherical flow
or a jet at $t<t_j$, when the edge of the jet is not visible, $\int
d\Omega=\int_0^{2\pi}d\phi\int_0^{y_{\rm max}}dy$, $U=0$ and
$\theta_p=\pi/2$. For $\epsilon=2$ and $y_{\rm max}=1$ we obtain
$-Q/IP_{\rm max}\equiv
f(\alpha)=\frac{(2^{4+\alpha}-1)(2+\alpha)(3+\alpha)}
{2^{7+\alpha}-28+2\alpha[2^{3+\alpha}5-7+(2^{3+\alpha}-1)\alpha]}$, so
$P_{\rm ord}/P_{\rm max}=f(\alpha)\approx 0.90-0.93$ for
$0<\alpha<1.5$. For $\alpha=\frac{p-1}{2}$, $P_{\rm
  max}=\frac{p+1}{p+7/3}$, $\epsilon=\frac{p+1}{2}$ (i.e. PLS G in
Granot \& Sari 2002), our analytic result corresponds to $p=3$, for
which $P_{\rm ord}/P_{\rm max}=\frac{93}{101}\approx 0.92$ and $P_{\rm
  ord}=\frac{279}{404}\approx 0.69$.  For $y_{\rm max}\gg 1$,
$\epsilon=2$ we obtain
$f(\alpha)=\frac{(2+\alpha)(3+\alpha)}{8+5\alpha+\alpha^2}$; for
$\alpha=\frac{p-1}{2}$ and $p=3$, $P_{\rm ord}/P_{\rm
  max}=\frac{6}{7}\approx 0.86$ and $P=\frac{9}{14}\approx 0.64$.  The
diference between the $y_{\rm max} = 1$ and $y_{\rm max}\gg 1$ results
may be relevant to the prompt GRB (see \S~\ref{discussion}), where the
tail of a pulse (which corresponds to $y>1$) is predicted to be less
polarized than its peak ($y\lesssim 1$). When the edge of the jet is
visible, the limits of integration over $d\Omega$ change. As this
causes relatively small modifications in $\theta_p$ and $P_{\rm ord}$,
we use the analytic expressions above for simplicity.

\begin{figure}
\plotone{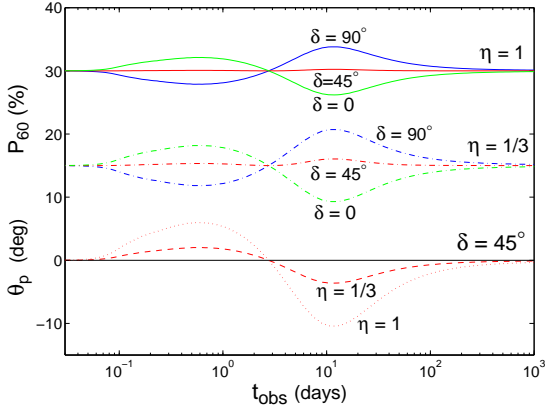}
\figcaption[]
{Polarization lightcurves for ordered $+$ random magnetic field
 components, for $b=0$ and the same jet parameters as in
 Fig. \ref{fig1}. As 
$\theta_p(\delta=0,90^\circ)\equiv 0$, we show only
$\theta_p(t,\delta=45^\circ)$.
\label{fig2}}
\vspace{-0.5cm}
\end{figure}

Figure \ref{fig2} depicts a sample of polarization lightcurves in
which both $\eta$ and $b$ are taken to be independent of time.  In
this case the $B_{\rm ord}$-induced polarization is constant (in both
$P$ and $\theta_p$) throughout the afterglow. Interestingly, a similar
polarization signature could be produced by dust in our galaxy or in
the GRB host galaxy. In the latter case, however, the polarization
would likely be accompanied by absorption that would redden the
spectrum: this could in principle make it possible to estimate the
level of the galactic dust contribution and thereby determine the
fraction of such a constant-polarization component that is intrinsic
to the source.  Since $P_{\rm ord}\approx P_{\rm max}$ whereas $P_{\rm
  rnd}$ is typically much smaller, we find that for 
$\eta=1$, and even for $\eta=1/3$, the
polarized intensity is still dominated by $B_{\rm ord}$ ($\eta P_{\rm
  ord}>P_{\rm rnd}$), with $B_{\rm rnd}$ only inducing relatively
small fluctuations around the $B_{\rm ord}$-induced values of $P$ and
$\theta_p$. For $\delta=45^\circ$ the fluctuations in both $P$ and
$\theta_p$ are very small in this parameter range.

If $B_{\rm ord}$ dominates the polarization, then, by equation
(\ref{P_tot}), the time evolution of $P$ follows that of the intensity
ratio $\eta$. The low measured values of $P$ indicate that $\eta \ll
1$, so $B_{\rm rnd}$ dominates the emissivity. To the extent that the
random field is close to equipartition ($\epsilon_{B,{\rm rnd}}\sim
1$), $\eta=\epsilon_{B,{\rm ord}}/\epsilon_{B,{\rm rnd}}\sim
\epsilon_{B,{\rm ord}}$. If the shock is radiative during its early
evolution, then cooling-induced compression increases the
emissivity-weighted $\epsilon_{B,{\rm ord}}$ over its immediate
postshock (adiabatic) value by a factor\footnote{This assumes that the
  fraction $\epsilon_e$ of the internal energy just behind the shock
  transition that resides in relativistic electrons and $e^\pm$ pairs
  is radiated away (Granot \& K\"onigl 2001).}
$\sim(1-\epsilon_e)^{-1}$.  The transition from fast to slow cooling,
which occurs at $t=t_0$, could therefore reduce $\eta$ and may
contribute to the early decline of $P$ observed in some sources.
During the subsequent, slow-cooling phase, $\epsilon_{B,{\rm ord}}$ is
essentially equal to the magnetization parameter of the ambient
medium, $\sigma=B_{\rm ext}^2/4\pi\rho_{\rm ext}c^2$, so the evolution
of $P$ during that phase may reflect the radial behavior of this
parameter: $\sigma$ is expected to be roughly constant for an ISM or a
stellar wind but to increase with $r$ inside a PWB (K\"onigl \& Granot
2002). If the orientation of the ambient field also changed with
radius then this would lead to a gradual variation in $\theta_p$. If
one approximates $B_{\rm ext}\propto r^{a/2}$ and $\rho_{\rm
  ext}\propto r^{-k}$, then $\epsilon_{B,{\rm ord}}\propto t_{\rm
  obs}^{\frac{a+k}{4-k}}$. We parameterize the above effects by
$\epsilon_{B,{\rm ord}}=\epsilon_{B,0} (t_{\rm
  obs}/t_0)^{\frac{a+k}{4-k}} F(t_{\rm obs}/t_0)$, where
$F(x)=1+\frac{g-1}{g+1}\frac{2}{\pi}\arctan(\xi\ln x)$ describes the
amplitude ($g$) and sharpness ($\xi$) of the change in
$\epsilon_{B,{\rm ord}}$ at $t_{\rm obs}\sim t_0$.  We assume that
$\epsilon_{B,{\rm tot}}=\epsilon_{B,{\rm ord}}+ \epsilon_{B,{\rm
    rnd}}={\rm const}$, so $\eta= (\frac{\epsilon_{B,{\rm
    tot}}}{\epsilon_{B,{\rm ord}}}-1)^{-1}$.

\begin{figure}
\plotone{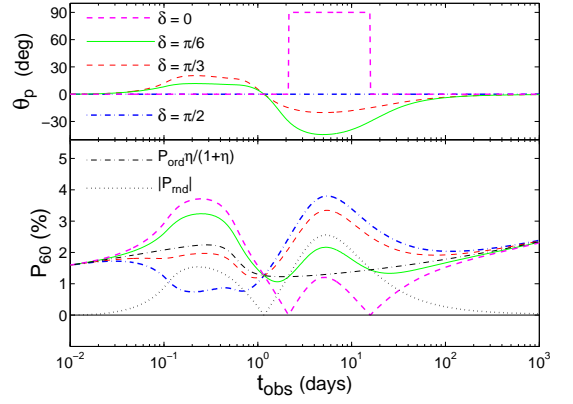}
\figcaption[]
{Polarization lightcurves for a jet with
$\theta_0=3^\circ$, $\theta_{\rm obs}=0.75\theta_0$, $b=0.5$,
$\epsilon_{B,{\rm tot}}=0.05$, $\epsilon_e=0.1$, $t_j\approx
t_0=14\;{\rm hr}$, and $\eta(t)$ given by
$\epsilon_{B,0}=0.0015$, $a=2$, $k=0$, $g=3$, and $\xi=3$.
\label{fig3}}
\vspace{-0.5cm}
\end{figure}

Figure \ref{fig3} shows an example of the polarization lightcurves.
The choice of parameters for $\eta(t)$ is motivated by GRB 020813, in
which $\theta_p$ was again roughly constant with time whereas $P$
first decreased (from $P \approx 2\%$ after $\sim 6\, {\rm hr}$ to
$P\approx 0.6\%$ after $\sim 24\, {\rm hr}$) and subsequently
increased monotonically, reaching $P \approx 3.7\%$ after $96\, {\rm
  hr}$. A sharp break in the lightcurve was observed after $14\;{\rm
  hr}$ (Covino et al. 2003b). It is seen that roughly equal
contributions to the polarization from $B_{\rm ord}$ and $B_{\rm rnd}$
can provide a qualitative fit to the evolution of $P$ and $\theta_p$
in this source for $\delta\approx 60^\circ-90^\circ$.  More generally,
the polarization lightcurves can show a diverse behavior that varies
as a function of $\delta$ as well as of $b$ and $\theta_{\rm
  obs}/\theta_0$. So long as $\eta P_{\rm ord}>P_{\rm rnd}$, the
changes in $\theta_p$ would be small whereas the variations in $P$
could be significant, as found observationally.

\section{Discussion}
\label{discussion}

The linear polarization in GRB afterglows may be largely due to an
ordered magnetic field in the ambient medium, which gives rise to an
ordered field component behind the afterglow shock that is coherent
over the entire emission region. This can result in a polarization
position angle (PA), $\theta_p$, that is roughly constant in time as
well as in a variable degree of polarization, $P$, as found in all
afterglow observations to date (except one; see footnote
\ref{021004}).

The magnetic field in the GRB ejecta is potentially much more ordered
than in the shocked ambient medium behind the afterglow shock,
reflecting the likely presence of a dynamically important,
predominantly tansverse, large-scale field advected from the source
(e.g., Spruit, Daigne, \& Drenkhahn 2001; Vlahakis \& K\"onigl 2001).
This could result in a large value of $P$ [up to $\sim
(0.90-0.93)P_{\rm max}\sim 60\%$] in the prompt $\gamma$-ray
emission\footnote{After this paper was submitted, Coburn \& Boggs
  (2003) reported a measurement of $P=80\pm 20\%$ in the $\gamma$-ray
  emission of GRB 021206, which is naturally (and most likely; Granot
  2003) produced in this way.}  as well as in the `optical flash' and
`radio flare', which are attributed to emission from the reverse
shock. If the polarization from the reverse shock is indeed dominated
by the ordered component, and if it is coherent over the whole ejecta,
then $\theta_p$ is not expected to vary significantly during the
optical flash or between the optical flash and the radio flare.
However, if the ordered magnetic field is coherent only in patches of
angular size $\theta_B$, then, so long as $\gamma>1/\theta_B$, we
expect $P\sim P_{\rm max}$, whereas after $\gamma$ drops below
$1/\theta_B$ we expect $P\sim \gamma\theta_BP_{\rm max}$ and
variations in $\theta_p$ on timescales $\Delta t_{\rm obs}\lesssim
t_{\rm obs}$ on account of the averaging over
$N\sim(\gamma\theta_B)^{-2}$ mutually incoherent patches within the
observed region of angle $1/\gamma$ about the line of sight. (This
resembles the proposal by Gruzinov \& Waxman 1999, except that here
$N$ is envisioned to increase with time.) In the latter case $P$ might
be smaller and $\theta_p$ would be different in the `radio flare' (for
which typically $\gamma\lesssim 10$) than in the `optical flash' (for
which $\gamma\gtrsim 100$).

Variability in the afterglow lightcurve, as reported in GRBs 021004
and 030329, whether induced by a clumpy external medium or a patchy
shell (Lazzati et al. 2002; Nakar, Piran, \& Granot 2002), should give
a different weight to emission from different parts of the afterglow
image, thus breaking its symmetry and inducing
polarization.\footnote{If the density distribution is spherically
  symmetric, the symmetry would need to be broken by the outflow
  geometry --- e.g., a jet observed off axis.}  Therefore, we expect a
highly variable lightcurve to be accompanied by variability in both
$P$ and $\theta_p$.\footnote{\label{021004}After this paper was
  submitted, a change of $45^\circ$ in $\theta_p$ was reported in GRB
  021004 between $9$ and $16\;$hr (Rol et al. 2003). This cannot be
  explained by simple jet models (Sari 1999; Ghisellini \& Lazzati
  1999) but could naturally arise in conjunction with the variability
  in the lightcurve (which, in fact, peaked at about the same time).}

Early polarization measurements, starting at $t_{\rm obs}\ll t_{j}$,
are crucial for distinguishing between our model and purely tangled
jet field models, as the latter predict $P(t_{\rm obs}\ll t_{j})\ll
P(t_{\rm obs}\sim t_j)$, whereas our model allows $P(t_{\rm obs}\ll
t_{j})\sim P(t_{\rm obs}\sim t_j)$.  In the latter models $P$ is
expected to peak, or else vanish and reappear rotated by $90^\circ$,
around $t_j$. In contrast, in our model, if the polarization is
dominated by an ordered magnetic field, then the variations in the
polarization around $t_j$ would be much less pronounced, with
$\theta_p$ exhibiting only a gradual variation and $P$ never crossing
zero. Our model predicts a possible change in $P$ around the
transition time from fast to slow cooling, $t_0$, where typically
$t_0\sim 1\;$hr ($\sim 1\;$day) for ISM-like (stellar wind-like)
parameters (although it may vary considerably around these values).

\vspace{-0.5cm}
\acknowledgments We thank P. Goldreich, R. Sari, A. Panaitescu, and
E. Rossi for useful discussions. This research was supported in
part by funds for natural sciences at the Institute for Advanced
Study (JG) and by NASA ATP grant NAG5-12635 (AK).

\end{document}